\pdfoutput=1
\documentclass[journal=jacsat,manuscript=preprint]{achemso}
\usepackage{graphicx}
\usepackage{siunitx}
\DeclareSIUnit\angstrom{\text{\AA}}
\usepackage{bm} 
\usepackage{chemformula} 
\usepackage[T1]{fontenc} 
\usepackage[hidelinks]{hyperref}

\author{Yu Miyazaki}
\affiliation[PFN]{Preferred Networks, Inc., 100-0004, 1-6-1 Otemachi, Chiyoda-ku, Tokyo, Japan.}
\email{yumiyazaki@preferred.jp}
\author{Atsuhiro Tomita}
\affiliation[PFN]{Preferred Networks, Inc., 100-0004, 1-6-1 Otemachi, Chiyoda-ku, Tokyo, Japan.}
\author{Akihide Hayashi}
\affiliation[PFN]{Preferred Networks, Inc., 100-0004, 1-6-1 Otemachi, Chiyoda-ku, Tokyo, Japan.}
\author{So Takamoto}
\affiliation[PFN]{Preferred Networks, Inc., 100-0004, 1-6-1 Otemachi, Chiyoda-ku, Tokyo, Japan.}
\author{Mizuki Takemoto}
\affiliation[PFN]{Preferred Networks, Inc., 100-0004, 1-6-1 Otemachi, Chiyoda-ku, Tokyo, Japan.}
\author{Hodaka Mori}
\affiliation[PFN]{Preferred Networks, Inc., 100-0004, 1-6-1 Otemachi, Chiyoda-ku, Tokyo, Japan.}

\title[PFPMM]{PFP/MM: A Hybrid Approach Combining a Universal Neural Network Potential with Classical Force Fields for Large-Scale Reactive Simulations}

\abbreviations{Neural Network Potentials}

\begin{document}

\begin{abstract}
Universal machine-learning interatomic potentials (uMLIPs) enable reactive molecular simulations with near-DFT accuracy, yet applying them efficiently to large, realistic condensed-phase systems remains computationally demanding.
Here we present PFP/MM, a hybrid approach that combines a uMLIP, PreFerred Potential (PFP), with molecular mechanics (MM) to enable both large-scale and long-time simulations that are challenging for uMLIP-only calculations.
Using an alanine dipeptide in explicit water, we achieve multi-ns/day enhanced sampling and obtain a Ramachandran plot consistent with established basins.
For an intramolecular nucleophilic addition reaction in a polar solvent environment, we reproduce the expected solvent-induced stabilization in the free-energy profile.
We further apply the approach to a cytochrome P450 Compound~I hydroxylation reaction and obtain a free-energy landscape consistent with the accepted reaction mechanism.
These results demonstrate that uMLIP-based reactive simulations can be applied to diverse condensed-phase processes in large, realistic environments.
\end{abstract}

\section{Introduction}
One practically important challenge in molecular simulation is to model chemical reactions in condensed-phase environments such as solutions, interfaces, and enzymes, which often requires reaction-path calculations and rare-event sampling\cite{Piccini2022-ab}. 
Detailed atomistic simulations of such processes can provide mechanistic insight and help guide the design of new chemical transformations. 
While conventional fixed-topology molecular mechanics (MM) force fields are highly efficient for long-timescale sampling in large systems, they are not inherently designed to describe bond breaking and bond forming or changes in electronic structure, for example charge redistribution or changes in oxidation state at metal centers\cite{Senftle2016-zn}. 
Therefore, quantum mechanical (QM) or quantum chemical methods are often required.
However, first-principles approaches such as Kohn-Sham density functional theory (DFT)\cite{Hohenberg1964-mm,Kohn1965-ss} exhibit steep computational scaling, and reactions strongly coupled to environmental fluctuations typically demand extensive conformational sampling\cite{Bowler2012-oq,Lin2019-qr}. 
As a result, routine applications to large condensed-phase systems remain challenging.

To reduce the computational cost, multiresolution approaches such as QM/MM\cite{Warshel1976-mw,Singh1986-hx,Field1990-ac}, where a QM Hamiltonian is applied to the region of interest such as the reactive center and an MM force field describing the surrounding environment, have long been proposed. 
Even in QM/MM, however, long-time sampling for rare events is often limited by the cost of the QM region, and the overall expense can remain prohibitive\cite{Vennelakanti2022-harder,Gopal2026-lr,Pultar2025-ir}. 
Recently, rapid progress in machine-learning interatomic potentials (MLIPs)\cite{Behler2007-tn,Bartok2010-qe,Tompson2015-qf}, which provide surrogate models for QM energies and forces, has motivated ML/MM (or NNP/MM) approaches \cite{Grassano2025-la,Galvelis2023-um,Gopal2026-lr,Pultar2025-ir,Karwounopoulos2025-mb,Ding2024-ut,Xu2021-da} that replace the QM component in QM/MM with an MLIP. 
Many existing ML/MM frameworks, however, rely on neural network potentials trained primarily for small organic molecules. 
For example, the ML/MM framework by Galvelis and co-workers\cite{Galvelis2023-um} employs ANI-2x\cite{Devereux2020-dx} and TorchMD-NET\cite{Pelaez2024-iu} potentials, and explicitly notes that ANI-2x is limited to H, C, N, O, F, S, and Cl. 
Such element restrictions hinder direct application, without additional MLIP retraining, to catalysis and metalloenzyme chemistry where inorganic elements and metal-centered reactivity are common.

Recent developments have begun to enable universal MLIPs (uMLIPs)\cite{Batatia2025-a,Chen2022-sn,Deng2023-rc,Takamoto2022-da} aimed at covering broad chemical space, reducing the need to construct or fine-tune a bespoke potential for each new target system. 
PreFerred Potential (PFP)\cite{Takamoto2022-da} is one such uMLIP trained on diverse DFT datasets, with an extended coverage up to 96 elements\cite{Shinagawa2026-PFP,Jacobs2025-sv}. 
This broad coverage provides a basis for treating a wide range of systems, including organic molecules, inorganic solids, and organic and inorganic hybrid materials, for polymers\cite{Mori2025-pfpoly,Taborosi2025-molecular,Mieda2025-bk}, interfaces\cite{Joutsuka2025-computational,Hu2025-hydroxylation}, and metal--organic frameworks\cite{Shimada2024-is,Koh2024-defect}.

In this work, we propose PFP/MM, a region-partitioned hybrid framework that combines the broad chemical coverage of the uMLIP, PFP with the efficiency of classical MM. 
In PFP/MM, only the chemically active region, optionally including a local portion of the surrounding environment when necessary, is described by PFP, while the remaining majority of atoms are treated with MM. 
This strategy enables large-scale reactive simulations and related computational tasks that are impractical with full-system uMLIP calculations. 
PFP/MM is implemented using OpenMM\cite{Eastman2024-om} as the MD engine, allowing GPU-accelerated evaluation of the MM region and making large hybrid simulations computationally practical. 
We first benchmark the speed and scalability of PFP/MM and demonstrate long-timescale simulations of alanine dipeptide in explicit solvent. 
Next, we show that the framework captures solvent-dependent stabilization in an intramolecular nucleophilic addition reaction in solution. 
Finally, we apply PFP/MM to a metalloenzyme reaction, the hydroxylation reaction mediated by cytochrome P450 Compound I, to demonstrate the feasibility of reactive simulations in a large, realistic biomolecular environment while leveraging the 96-element coverage of PFP. 
This example highlights the practical advantage of uMLIP chemical coverage for metal-centered reactivity in biomolecular environments, which can be difficult to address using ML/MM approaches with limited element sets.

\section{Method}
PFP/MM follows the standard additive hybrid scheme used in QM/MM (or ML/MM) approaches.\cite{Senn2009-qm,Galvelis2023-um}
Let the potential energy of the MLIP region be $U_{\mathrm{ML}}(\mathbf{r}_{\mathrm{ML}})$, that of the MM region be $U_{\mathrm{MM}}(\mathbf{r}_{\mathrm{MM}})$, and the interaction between the MLIP and MM regions be $U_{\mathrm{ML-MM}}(\mathbf{r}_{\mathrm{ML}},\mathbf{r}_{\mathrm{MM}})$.
The total hybrid potential is given by
\begin{equation}\label{eq:addition}
  U_{\mathrm{ML/MM}}(\mathbf{r}_{\mathrm{ML}},\mathbf{r}_{\mathrm{MM}})
  =U_{\mathrm{ML}}(\mathbf{r}_{\mathrm{ML}})
  +U_{\mathrm{MM}}(\mathbf{r}_{\mathrm{MM}})
  +U_{\mathrm{ML-MM}}(\mathbf{r}_{\mathrm{ML}},\mathbf{r}_{\mathrm{MM}}).
\end{equation}

We adopt mechanical embedding for the interfacial coupling, where the MLIP--MM interaction is described by classical nonbonded interactions:
\begin{equation*}
  U_{\mathrm{ML-MM}}(\mathbf{r}_{\mathrm{ML}},\mathbf{r}_{\mathrm{MM}})
  = \sum_{i\in \mathrm{ML}}\sum_{j\in \mathrm{MM}}
  \left[
    \frac{q_iq_j}{\left|\mathbf{r}_i-\mathbf{r}_j\right|}
    +4\epsilon_{ij}\left\{
      \left(\frac{\sigma_{ij}}{\left|\mathbf{r}_i-\mathbf{r}_j\right|}\right)^{12}
      -\left(\frac{\sigma_{ij}}{\left|\mathbf{r}_i-\mathbf{r}_j\right|}\right)^{6}
    \right\}
  \right],
\end{equation*}
where $q_i$ is the partial charge of atom $i$, and $\epsilon_{ij}$ and $\sigma_{ij}$ are the Lennard--Jones parameters between atoms $i$ and $j$.

When a covalent bond crosses the boundary between the MLIP (PFP) and MM regions, naively truncating the system creates dangling bonds in the MLIP region, which can deteriorate the accuracy of the MLIP inference.
To mitigate this issue, the ML/MM scheme caps each dangling bond in the MLIP region with a virtual hydrogen atom (a link atom) during the MLIP evaluation.\cite{Singh1986-hx}
See Supporting Information for the detailed definition, including the force redistribution procedure.

In the implementation of PFP/MM, PFP was used as the MLIP (i.~e. $U_{\mathrm{ML}}(\mathbf{r}_{\mathrm{ML}})\equiv U_{\mathrm{PFP}}(\mathbf{r}_{\mathrm{ML}},\mathbf{Z}_{\mathrm{ML}})$, where $\mathbf{Z}_{\mathrm{ML}}$ is the atomic numbers in the MLIP region).
The architecture of PFP is based on TeaNet,\cite{Takamoto2022-xc} which incorporates a second-order Euclidean tensor into the graph neural network and maintains the necessary equivariances.
All PFP calculations in this paper used PFP \texttt{v6.0.0}.
In this work, we used two modes of PFP \texttt{v6.0.0}, \texttt{CRYSTAL\_U0\_PLUS\_D3} and \texttt{MOLECULE}.
The \texttt{CRYSTAL\_U0\_PLUS\_D3} mode is trained on a wide range of structures across 96 elements using the Perdew--Burke--Ernzerhof (PBE) functional\cite{Perdew1996-qh} with the projector-augmented wave (PAW) method\cite{Blochl1994-paw} and plane-wave basis, including Grimme's D3 dispersion correction.\cite{Grimme2010-df}
The \texttt{MOLECULE} mode is trained on small organic molecules using the $\omega$B97X-D functional\cite{Chai2008-bu} with the 6-31G(d) basis set.\cite{Ditchfield1971-gf}
In addition to an NVIDIA GPU implementation, PFP is also available with a backend for MN-Core~2,\cite{Namura2021-mncore,MNCore2WhitePaperEN2023} an in-house AI accelerator developed by Preferred Networks, Inc. that is optimized for dense matrix-multiplication workload.

The implementation of PFP/MM is based on OpenMM,\cite{Eastman2024-om} a widely used GPU-accelerated MD engine.
PFP/MM consists of two main components:
(i) a routine that constructs an OpenMM \texttt{System} object from the full system parameterized with a classical force field, in which the terms assigned to the PFP region are removed; and
(ii) a routine that creates an OpenMM \texttt{Force} object that evaluates the potential energy (and forces) of the subsystem selected as the PFP region.
By adding the \texttt{Force} generated in (ii) to the \texttt{System} constructed in (i), we realize the additive ML/MM formulation in Eq.~\ref{eq:addition}.

\section{Results and Discussion}
\subsection{Benchmark of Alanine Dipeptide in Water}
\begin{figure}[p]
    \centering
    \includegraphics{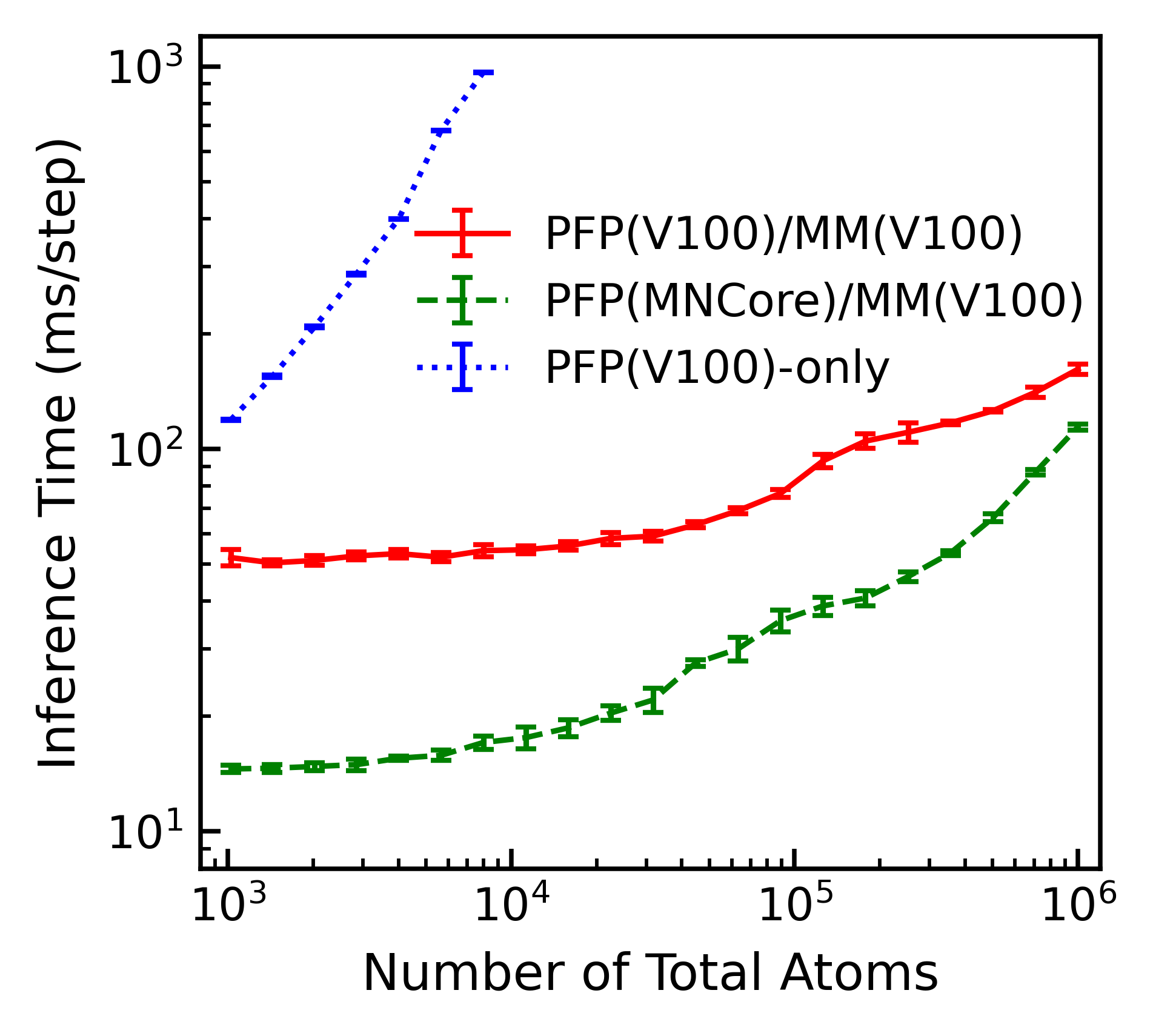}
    \caption{Speed benchmark in alanine dipeptide in water system.
In PFP/MM, the 22 atoms of the alanine dipeptide molecule are treated as the PFP region, and all remaining water molecules are treated as the MM region.
In PFP-only, all atoms are treated within PFP. 
Error bars represent standard errors over 5 runs.}
    \label{fig:speed}
\end{figure}

\begin{figure*}[p]
    \centering
    \includegraphics[width=\linewidth]{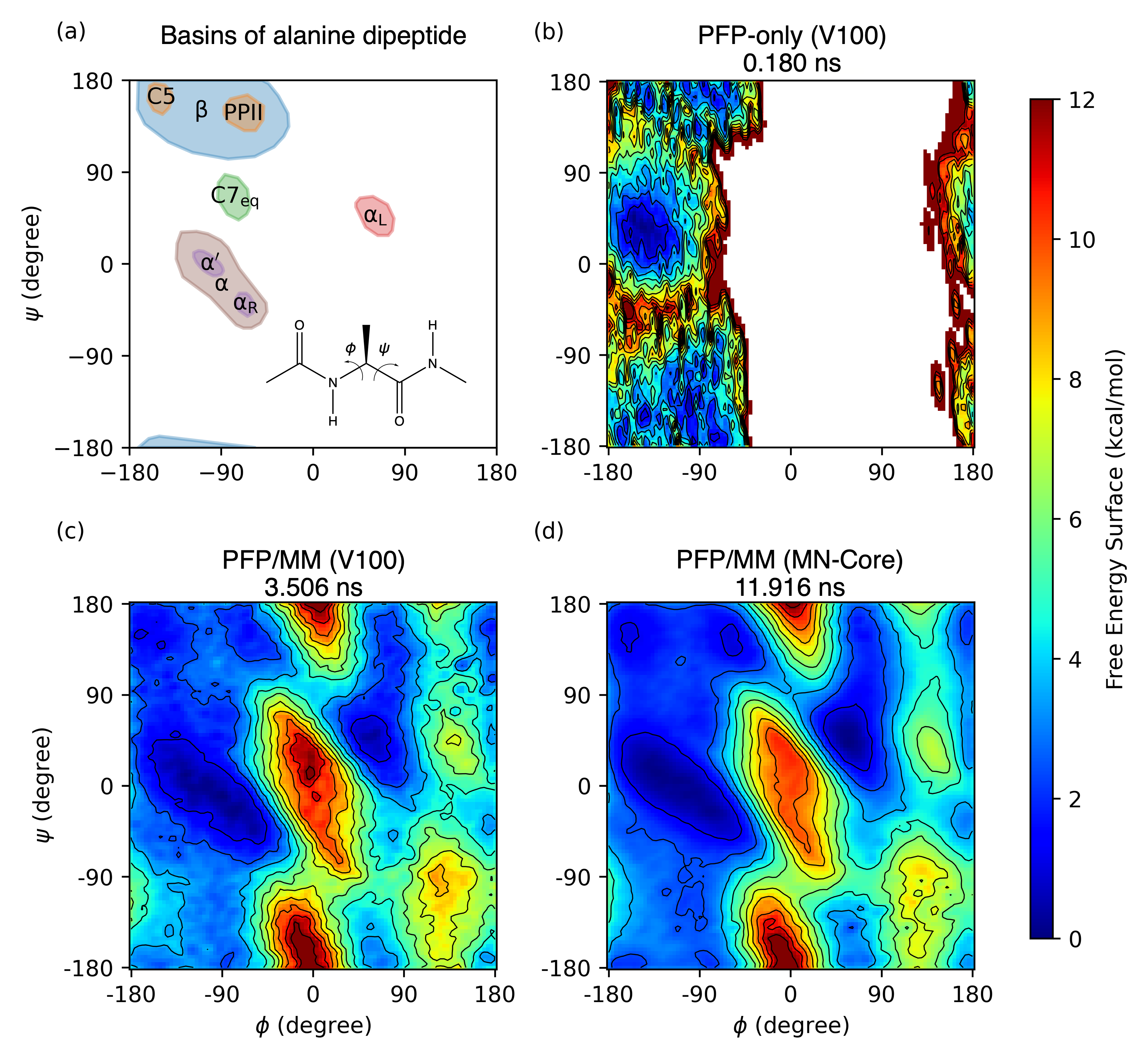}
    \caption{(a) Schematic overview of major conformational basins of the alanine dipeptide in water. 
    Inset shows the definition of the backbone dihedral angles $\phi$ and $\psi$.
Adapted from the original work of Feig \cite{Feig2008-of}.
(b--d) Free energy surfaces from one-day simulations: (b) PFP-only, (c) PFP/MM with V100, and (d) PFP/MM with MN-Core~2.}
    \label{fig:ala2-oneday}
\end{figure*}
To assess the performance of PFP/MM relative to PFP-only, we carried out benchmarks on an alanine dipeptide-in-water system.
Figure~\ref{fig:speed} shows the inference time per MD step as a function of the total number of atoms.
In PFP/MM, the solute (one alanine dipeptide molecule; 22 atoms) defines the PFP region, whereas all water molecules are treated in the MM region.
All MM calculations were performed on an NVIDIA Tesla V100 GPU, while inference for the PFP region was executed either on a V100 or on the MN-Core~2 accelerator. In the PFP-only setup, all atoms were treated by PFP and the simulation was performed on a V100.
Under the present settings, the largest system that could be simulated with PFP-only contained 7,975 atoms, whereas PFP/MM can, in principle, handle systems exceeding one million atoms.
For the 7,975-atom system, PFP/MM was 17.8-fold faster on a V100 and 56.5-fold faster on MN-Core~2 than PFP-only.
When the PFP region is sufficiently small, MN-Core~2 provides an additional speedup over V100 for the PFP inference.
For systems with fewer than $10^4$ MM atoms, the overall runtime is dominated by the PFP region and is nearly insensitive to the system size; above $10^4$ MM atoms, the MM cost becomes non-negligible and the runtime increases with the number of atoms, reducing the performance gap between V100 and MN-Core 2.
For a 1,000,002-atom system, PFP/MM achieved $\SI{0.161}{\second/step}$ (V100) and $\SI{0.114}{\second/step}$ (MN-Core~2), corresponding to $\SI{1.07}{\nano\second\per day}$ and $\SI{1.51}{\nano\second\per day}$, respectively, with a time step of $\Delta t=\SI{2}{\femto\second}$.

To evaluate the practical impact on enhanced sampling, we performed one-day well-tempered metadynamics simulations\cite{Barducci2008} and computed the Ramachandran FES of alanine dipeptide in water (Figure~\ref{fig:ala2-oneday}).
The simulation conditions were identical to those used in the speed benchmark (7,975 atoms in total; 22 PFP atoms for PFP/MM).
Figure~\ref{fig:ala2-oneday}a summarizes the major conformational basins reported for alanine dipeptide in water.\cite{Vymetal2011-df} Because PFP-only reached $\SI{0.180}{\nano\second\per day}$, the resulting sampling was insufficient to converge the FES, and the region with $\phi>\SI{0}{\degree}$ was scarcely visited (Figure~\ref{fig:ala2-oneday}b).
In contrast, PFP/MM enabled $\SI{3.506}{\nano\second\per day}$ on a V100 (Figure~\ref{fig:ala2-oneday}c) and $\SI{11.916}{\nano\second\per day}$ on MN-Core 2 (Figure~\ref{fig:ala2-oneday}d), allowing nanosecond-scale sampling within a single day.
Both PFP/MM runs reproduced the reported conformational basins, and the longer MN-Core 2 trajectory provided a more detailed FES.

\subsection{Intramolecular Nucleophilic Addition of an Amine to a Carbonyl: Solvent Effects}
\begin{figure*}[p]
    \centering
    \includegraphics[width=\linewidth]{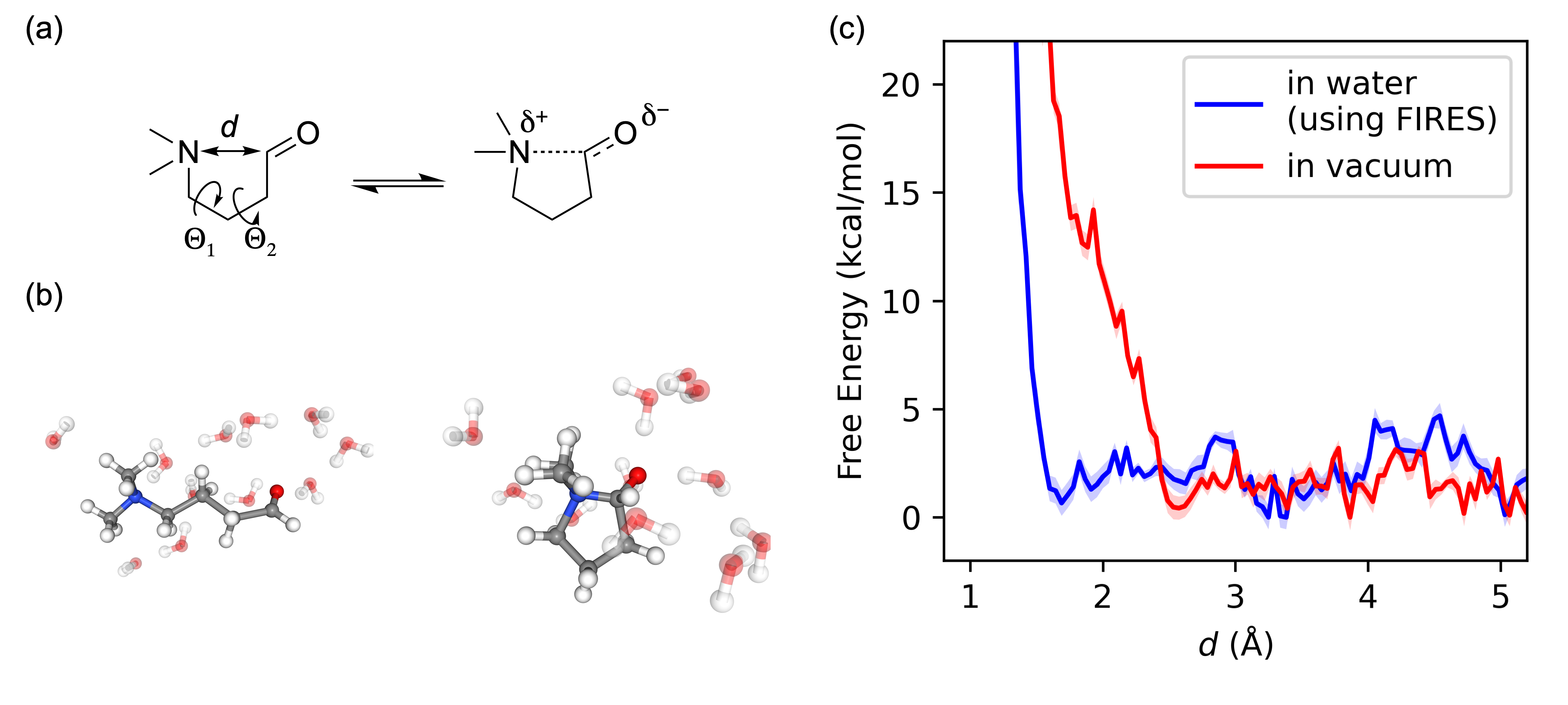}
    \caption{(a) Skeletal formulas of the open and cyclized NCO molecules.
    (b) Snapshots of the PFP region in the open and cyclized conformations.
    (c) Free-energy profiles of NCO in water (using FIRES) and in vacuum. 
    Regions around the lines indicates standard error estimated by block averaging of 5 blocks.}
    \label{fig:nco}
\end{figure*}
In the present PFP/MM implementation, interactions across the PFP/MM boundary are treated via mechanical embedding\cite{Lin2007-ns}; therefore, the PFP region experiences only classical electrostatic interactions from the MM region.
As a consequence, reactions in which solvent polarity and hydrogen bonding play a decisive role may require that not only the solute, but also nearby solvent molecules, be included in the PFP region\cite{Rowley2012-ke,Gopal2026-lr}.
To incorporate such local solvent effects while minimizing perturbations to ensemble averages, we employed FIRES (Flexible Inner Region Ensemble Separator), which applies a flexible restraint to molecules in the vicinity of the solute.\cite{Rowley2012-ke}

We investigated an intramolecular nucleophilic addition in a molecule containing a carbonyl and an amine group (hereafter denoted NCO; Figure~\ref{fig:nco}a).
In polar solvents, the approach of the amine N toward the carbonyl C can lead to a ring-closed, zwitterion-like structure.\cite{Boereboom2018-gw,Pilme2007-unusual} Using PFP/MM with FIRES, we computed the free-energy surface of NCO in water via metadynamics.
As collective variables (CVs), we used (i) the distance $d$ between the carbonyl C and the amine N and (ii) a torsional measure of the five-membered ring defined as $\Theta=\sqrt{\Theta_1^2+\Theta_2^2}$ as shown in Figure~\ref{fig:nco}a following ref.~\citenum{Boereboom2018-gw}.
We ran $\SI{1.2}{\nano\second}$ of well-tempered metadynamics.
Figure~\ref{fig:nco}b shows representative snapshots of the PFP region in the open and cyclized states, respectively.
In the cyclized state, a water molecule forms a hydrogen bond to the carbonyl oxygen, which is expected to contribute to stabilization of the zwitterion-like structure.
Figure~S1 in the Supporting Information shows the relationship between the C--N distance $d$ and the carbonyl C--O distance, along with Bader charges inferred by PFP.
In the cyclized state (small $d$), the carbonyl double bond becomes more single-bond-like; correspondingly, the oxygen becomes more negatively charged and the nitrogen becomes more positively charged, consistent with the zwitterion-like character depicted in Figure~\ref{fig:nco}a.
Figure~\ref{fig:nco}c compares free-energy profiles (projected onto $d$) for NCO in water and in vacuum.
In water, the cyclized state is stabilized, whereas it is unstable in vacuum, highlighting the essential role of solvation.
Boereboom and co-workers evaluated the free energies of reactant, transition state, and product for the same reaction using QM/MM combined with Difference-based Adaptive Solvation (DAS).\cite{Boereboom2018-gw} Our observation that the cyclized state is stabilized in polar solvent is consistent with that work, although quantitative free-energy values differ.
Possible origins include the substantially longer simulation time used here ($\SI{1.2}{\nano\second}$ vs $\SI{125}{\pico\second}$) as well as methodological differences between DAS and FIRES.

\subsection{P450-Cpd-I-Mediated Hydroxylation Reaction}
\begin{figure*}[p]
    \includegraphics[width=\linewidth]{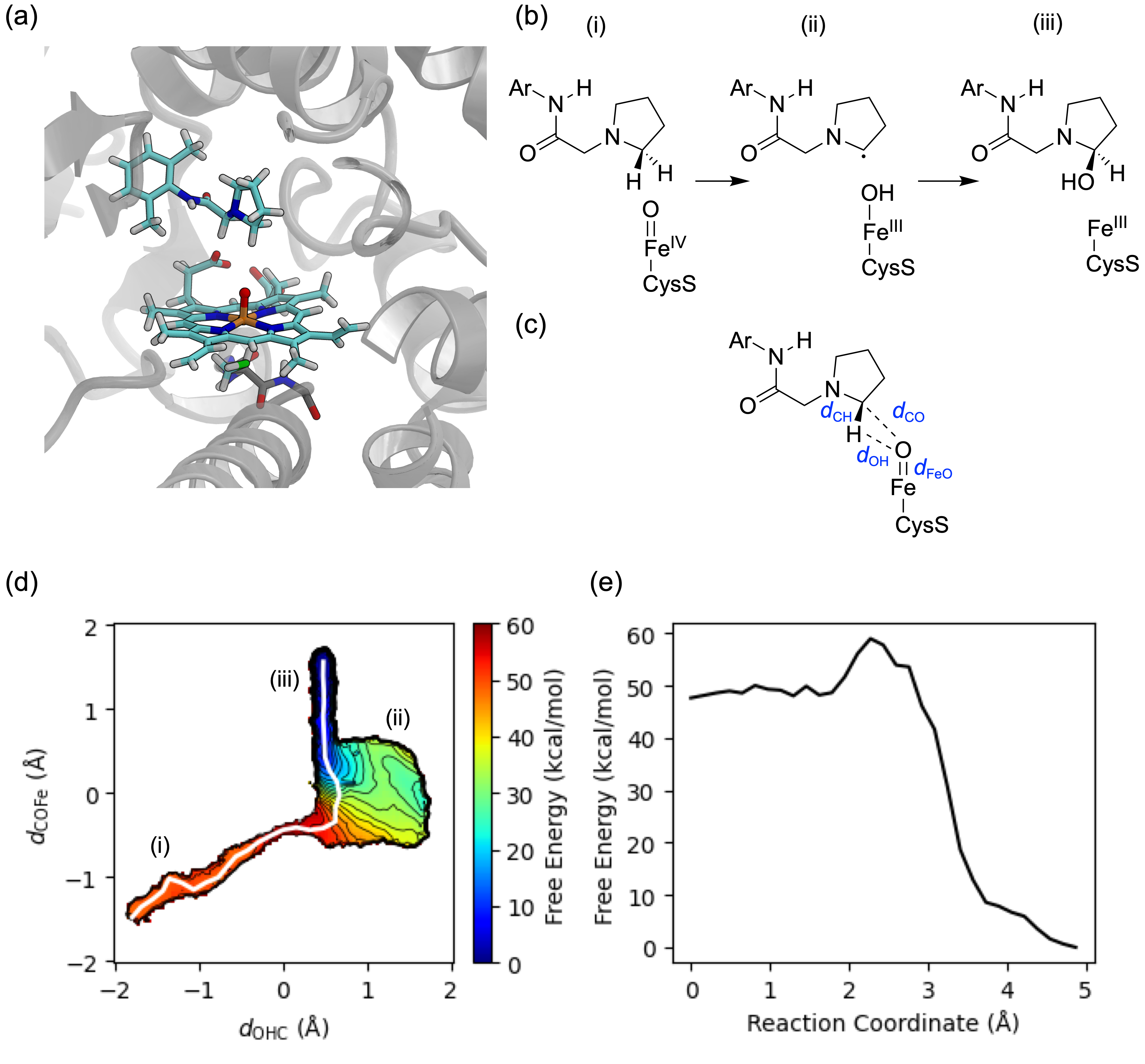}
    \caption{(a) Structure of the P450-Cpd-I enzyme-ligand model used in the PFP/MM simulation. 
    The PFP region is shown in color, while the MM region is shown in gray. 
    (b) Hydroxylation reaction mechanism of P450-Cpd-I proposed in ref.~\citenum{Wang2021-sb}.
    (c) Collective variables used in the metadynamics and umbrella sampling simulations.
    (d) Free energy surface obtained from umbrella sampling and the minimum free energy path. 
    (e) Free energy profile along the reaction coordinate.}
    \label{fig:P450}
\end{figure*}
As an example that leverages the universality of PFP, we studied a substrate hydroxylation reaction catalyzed by the metalloenzyme P450 Compound I (Cpd-I) shown in Figure~\ref{fig:P450}a.
The proposed mechanism (Figure~\ref{fig:P450}b) involves hydrogen atom abstraction (HAA) by the oxo ligand of Cpd-I, followed by rebound of the hydroxyl group to complete hydroxylation.\cite{Wang2021-sb,Prier2017-ps,Ren2016-id} 
The PFP region comprised the substrate, the porphyrin, and the cysteine ligand coordinating the Fe center, whereas the remainder of the protein and the aqueous solvent were treated in the MM region.
The covalent boundary was handled using the link-atom approach (see the Supporting Information).

We first performed pre-metadynamics to obtain a qualitative free-energy surface and representative structures (i), (ii), and (iii) (see Figure.~S2 in Supporting Information).
As collective variables, we used $d_{\mathrm{OHC}}=d_{\mathrm{CH}}-d_{\mathrm{OH}}$, which expresses whether the hydrogen atom is closer to the substrate C or to the oxo O, and $d_{\mathrm{COFe}}=d_{\mathrm{FeO}}-d_{\mathrm{CO}}$, which distinguishes O bound to Fe from O inserted into the substrate C (Figure~\ref{fig:P450}c).
To obtain a more quantitative landscape, we carried out umbrella sampling over 147 windows around the reactive region identified by metadynamics.
Each window was equilibrated for $\SI{100}{\pico\second}$ and sampled for $\SI{100}{\pico\second}$, for a total of $\SI{29.4}{\nano\second}$.
The FES was reconstructed using MBAR,\cite{Shirts2008-tn} and the minimum free-energy path was estimated using the string method (Figure~\ref{fig:P450}d).
The resulting FES indicates that HAA is the dominant free-energy barrier.
Along the minimum free-energy path, the system proceeds directly to (iii); even when the trajectory passes through (ii), the hydroxyl group rapidly migrates to yield (iii) without an additional barrier.
This behavior is consistent with previous QM/MM PES scans.\cite{Wang2021-sb} 
Figure~\ref{fig:P450}e shows the free-energy profile along the reaction coordinate, from which we estimate an activation free energy of $\sim\SI{12}{kcal/mol}$ for HAA and an overall free-energy difference of $\sim\SI{48}{kcal/mol}$ between the initial and final states.

It is possible to train a dedicated MLIP for a specific metal center\cite{Ding2024-ut,Xu2021-da}, but such approaches often require system- and element-specific model development, which limits their transferability across chemically diverse targets. 
Moreover, stable simulations frequently demand large QM training datasets, and the cost of generating the corresponding reference data can be substantial.

\section{Conclusion}
We presented PFP/MM, a hybrid approach that combines the universal MLIP PreFerred Potential (PFP) with classical molecular mechanics (MM) by treating chemically active regions with PFP and the remaining majority of atoms with a fast MM description.
This domain decomposition enables a favorable balance between accuracy and computational efficiency while retaining the broad transferability of a uMLIP.
Benchmarks on alanine dipeptide in water demonstrated substantial speedups over PFP-only (17.8-fold on V100 and 56.5-fold on MN-Core 2 for a 7,975-atom system) and confirmed that nanosecond-per-day throughput remains feasible even at the million-atom scale.
In one-day well-tempered metadynamics, PFP/MM achieved nanosecond-scale sampling and produced a stable Ramachandran free-energy surface reproducing the reported conformational basins, whereas PFP-only suffered from insufficient sampling.
Recognizing that mechanical embedding limits the ability of the MM environment to polarize the PFP region, we showed that solvent polarity and hydrogen bonding effects can be incorporated by explicitly including nearby solvent molecules in the PFP region with FIRES.
For an intramolecular nucleophilic addition, this strategy reproduced solvent-dependent stabilization of a zwitterion-like cyclized state in water relative to vacuum.
Finally, we applied PFP/MM to a metalloenzyme reaction (P450-Cpd-I-mediated hydroxylation) and obtained a free-energy landscape consistent with the accepted mechanism, in which hydrogen atom abstraction constitutes the primary barrier and hydroxyl rebound proceeds without an additional barrier.
Overall, PFP/MM provides a practical framework for large-scale, long-time reactive simulations that would be difficult to access with uMLIPs alone under realistic computational budgets.

\begin{acknowledgement}
We thank Mr. Ryosuke Kuwabara (MITSUI \& CO., LTD.), Dr. Kunio Sannohe (IBLC Co., Ltd.), Prof. Masayoshi Takayanagi (Shiga University), Dr. Masataka Yamauchi (Matlantis Corporation), Mr. Shunsuke Tonogai (Preferred Networks, Inc.), and Mr. Takahiro Okawara (Preferred Networks, Inc.)  for helpful discussions. 
We also thank Dr. Kaoru Hisama (Preferred Networks, Inc.) for pointing us to relevant literature.
We also thank Mr. Tomokazu Higuchi (Preferred Networks, Inc.) and Mr. Hiroya Kaneko (Preferred Networks, Inc.) for valuable comments about MN-Core~2.
\end{acknowledgement}

\bibliography{ref}

\end{document}


\renewcommand{\thefigure}{S\arabic{figure}}
\renewcommand{\thetable}{S\arabic{table}}

\begin{figure}[h]
    \centering
    \includegraphics[width=\linewidth]{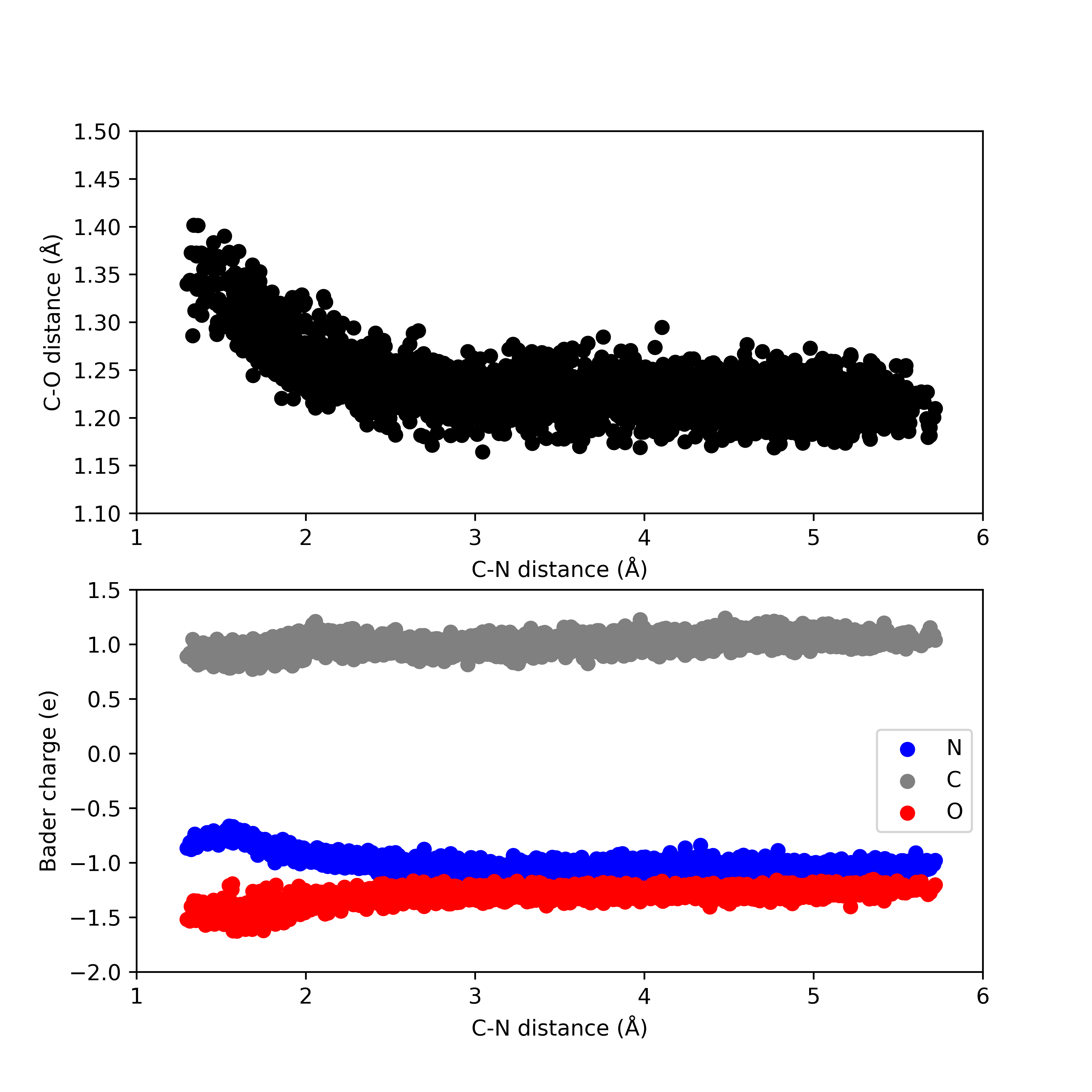}
    \caption{Scatter plots of the C--O distance (top) and Bader charge of N, C, and O atoms (bottom) as a function of the C--N distance obtained from metadynamics of NCO-water system.}
\end{figure}
\begin{figure}[h]
    \centering
    \includegraphics[width=\linewidth]{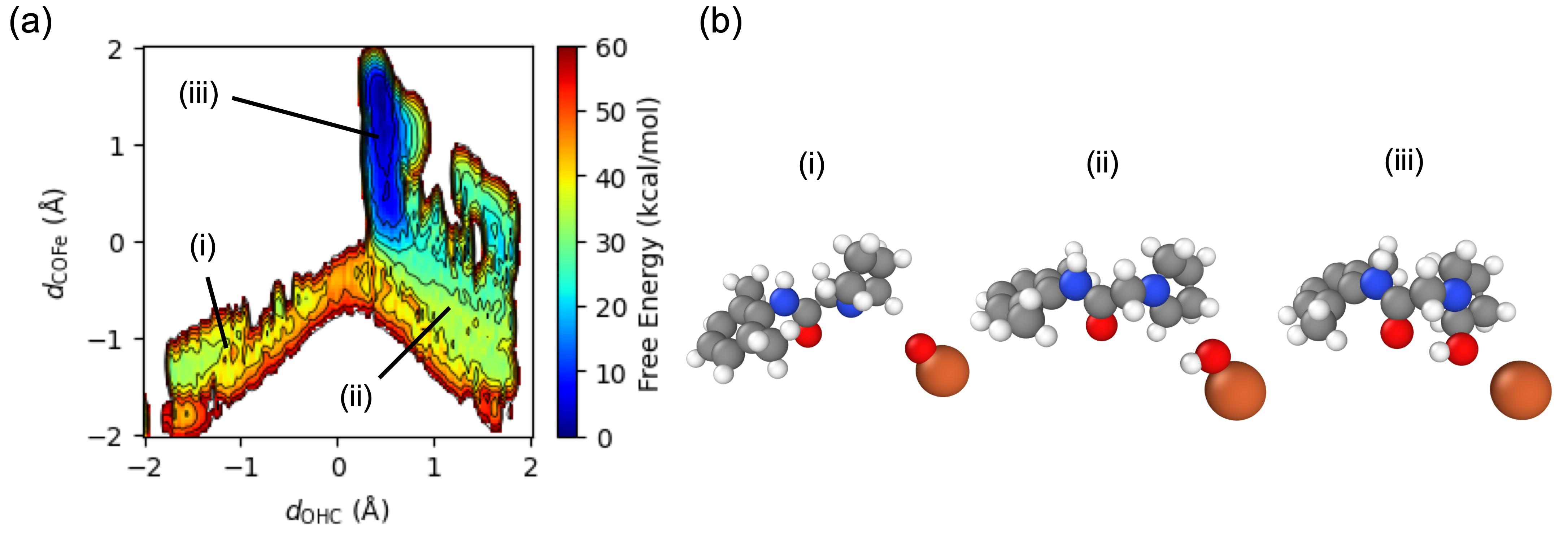}
    \caption{
      (a) Free energy surface obtained from pre-metadynamics and (b) the representative structures of (i), (ii), and (iii) shown in (a).
    }
\end{figure}

\section{Link Atom Treatment}
In PFP/MM, the system is partitioned into a PFP region and an MM region. 
If a covalent bond crosses the boundary, a naive partitioning leaves a dangling bond on the PFP side, which can destabilize the electronic structure and the resulting energy/forces. 
To mitigate this, we employ the link atom method\cite{Singh1986-hx}, where the severed bond on the PFP side is capped by a dummy atom (typically H) during the PFP calculation.

Because the PFP calculation is performed for a fictitious system consisting of the PFP region plus link atoms, PFP/MM must specify (i) how link atoms are placed and (ii) how the forces acting on them are handled.

\subsection{Placement of Link Atoms}
Consider a boundary bond between a PFP atom Q and an MM atom M, and denote the link atom by L. 
With position vectors $\mathbf{r}_{\mathrm{Q}}, \mathbf{r}_{\mathrm{M}}, \mathbf{r}_{\mathrm{L}}$, L is placed on the line segment QM as
\begin{equation}
  \mathbf{r}_{\mathrm{L}}=(1-C_{\mathrm{L}})\mathbf{r}_{\mathrm{Q}}+C_{\mathrm{L}}\mathbf{r}_{\mathrm{M}} .
\end{equation}
Two choices are commonly used for determining $C_{\mathrm{L}}$: (i) fixing the Q--L distance $|\mathbf{r}_{\mathrm{L}}-\mathbf{r}_{\mathrm{Q}}|$, or (ii) fixing the internal division ratio $C_{\mathrm{L}}$ itself\cite{Melo2018-jp}.
PFP/MM supports both schemes.
In this work, the link atom position was defined by fixing the C--H bond length to $\SI{1.09}{\angstrom}$.

\subsection{Redistribution of Link-atom Forces}
A PFP calculation returns forces for the fictitious PFP+link-atom system, including $\widetilde{\mathbf{F}}_{\mathrm{Q}}$ and the force on the link atom $\widetilde{\mathbf{F}}_{\mathrm{L}}$. 
Since L does not exist in the physical PFP/MM system, simply discarding $\widetilde{\mathbf{F}}_{\mathrm{L}}$ breaks translational and rotational invariance, leading to violations of conservation laws in molecular dynamics. 
Therefore, in PFP/MM we redistribute $\widetilde{\mathbf{F}}_{\mathrm{L}}$ onto the real atoms Q ($\widetilde{\mathbf{F}}_{\mathrm{Q}}$) and M ($\widetilde{\mathbf{F}}_{\mathrm{M}}$).

Let the unit vector along the Q--M bond be
\[
  \hat{\mathbf{e}}_{\mathrm{b}}=\frac{\mathbf{r}_{\mathrm{M}}-\mathbf{r}_{\mathrm{Q}}}{|\mathbf{r}_{\mathrm{M}}-\mathbf{r}_{\mathrm{Q}}|} .
\]
Then the redistributed contributions $\mathbf{F}'_{\mathrm{Q}}$ and $\mathbf{F}'_{\mathrm{M}}$ are given by
\begin{align}
  \mathbf{F}'_{\mathrm{Q}}&=(1-C_{\mathrm{L}})\widetilde{\mathbf{F}}_{\mathrm{L}}
  +C_{\mathrm{L}}\left(\widetilde{\mathbf{F}}_{\mathrm{L}}\!\cdot\!\hat{\mathbf{e}}_{\mathrm{b}}\right)\hat{\mathbf{e}}_{\mathrm{b}},\\
  \mathbf{F}'_{\mathrm{M}}&=C_{\mathrm{L}}\widetilde{\mathbf{F}}_{\mathrm{L}}
  -C_{\mathrm{L}}\left(\widetilde{\mathbf{F}}_{\mathrm{L}}\!\cdot\!\hat{\mathbf{e}}_{\mathrm{b}}\right)\hat{\mathbf{e}}_{\mathrm{b}} .
\end{align}
This construction (i) satisfies translational invariance through
$\mathbf{F}'_{\mathrm{Q}}+\mathbf{F}'_{\mathrm{M}}=\widetilde{\mathbf{F}}_{\mathrm{L}}$,
and (ii) satisfies rotational invariance through
$\left(\mathbf{r}_{\mathrm{L}}-\mathbf{r}_{\mathrm{Q}}\right)\times\widetilde{\mathbf{F}}_{\mathrm{L}}
=\left(\mathbf{r}_{\mathrm{M}}-\mathbf{r}_{\mathrm{Q}}\right)\times\mathbf{F}'_{\mathrm{M}}$.
In addition, since the Q--M bond potential is governed by the MM force field, we impose that (iii) the PFP+link-atom contribution does not drive bond stretching along $\hat{\mathbf{e}}_{\mathrm{b}}$.
That is $\left[\left(\mathbf{F}_{\mathrm{Q}}' + \widetilde{\mathbf{F}}_{\mathrm{Q}}\right) - \mathbf{F}_{\mathrm{M}}'\right] \cdot \hat{\mathbf{e}}_\mathrm{b} = 0$.
These three conditions uniquely give the above form.

\section{Computational Details}
Unless otherwise noted, electrostatic interactions were evaluated using the particle mesh Ewald (PME) method\cite{Darden1993-pme} with a \SI{1}{\nano\meter} real-space cutoff and an error tolerance of $ 5 \times 10^{-4}$. 
Temperature was controlled at $\SI{300}{\kelvin}$ with a Langevin thermostat, and pressure was maintained at $\SI{1}{bar}$ using a Monte Carlo barostat.
Well-tempered metadynamics and umbrella sampling simulations were performed using the PLUMED~2.8\cite{Tribello2014-mm,Podgorski2022-qd} plugin.
\subsection{Alanine Dipeptide in Water}
The MM region was described with the AMBER ff14SB force field\cite{Maier2015-ss}.
All bonds involving hydrogen atoms were constrained.
PFP calculations employed the \texttt{CRYSTAL\_U0\_PLUS\_D3} mode.
Parameters of Langevin thermostat were a timestep of $\Delta t=\SI{2}{\femto\second}$ and a friction coefficient of $\SI{1}{\per\pico\second}$.
After geometry optimization using the L-BFGS algorithm (energy tolerance: $\SI{10}{\kilo\joule\per\mole}\simeq \SI{2.4}{\kilo cal\per\mole}$), we performed a \SI{20}{\pico\second} MD run for equilibration, and the resulting structure was used as the initial configuration for the subsequent well-tempered metadynamics simulation. 
This optimization and equilibration procedure was not included in the reported computational time.
A well-tempered metadynamics bias was applied to the CVs $(\phi,\psi)$ with a deposition stride of 100 MD steps, a Gaussian height of $\SI{0.05}{\electronvolt}\simeq \SI{1.2}{kcal\per\mol}$, Gaussian widths $\sigma_{\phi}=0.2~\mathrm{rad},\sigma_{\psi}=0.2~\mathrm{rad}$, and a bias factor of $\gamma=10.0$.

\subsection{Intramolecular Nucleophilic Addition of an Amine to a Carbonyl}
The MM region was modeled with OpenFF 2.1.0\cite{Boothroyd2023-rr}, and partial charges were assigned using AM1-BCC\cite{Jakalian2000-qq,Jakalian2002-qd}.
The PFP region included eight water molecules surrounding the carbonyl oxygen and three water molecules near the amine nitrogen.
For these selected water molecules, FIRES restraints were applied to the oxygen atoms.
PFP calculations used the \texttt{MOLECULE} mode.
Parameters of Langevin thermostat were a timestep of $\Delta t=\SI{0.5}{\femto\second}$ and a friction coefficient of $\SI{2}{\per\pico\second}$.
After geometry optimization using the L-BFGS algorithm (energy tolerance: $\SI{10}{\kilo\joule\per\mole}\simeq \SI{2.4}{\kilo cal\per\mole}$), we performed a \SI{5}{\pico\second} MD run for equilibration, and the resulting structure was used as the initial configuration for the subsequent well-tempered metadynamics simulation. 
A well-tempered metadynamics bias was applied to the CVs $(d,\Theta)$ with a deposition stride of 100 MD steps, a Gaussian height of $\SI{0.05}{\electronvolt}\simeq \SI{1.2}{kcal\per\mol}$, Gaussian widths $\sigma_{d}=\SI{0.1}{\angstrom},\sigma_{\Theta}=0.1\pi~\mathrm{rad}$, and a bias factor of $\gamma=20.0$.

\subsection{P450-Cpd-I-Mediated Hydroxylation Reaction}
Standard amino-acid residues were described with the AMBER ff19SB force field\cite{Tian2020-it}, while the remaining regions were parameterized in the GAFF2 format\cite{Wang2004-tn}.
Parameters for nonstandard residues were derived using the MCPB.py tool\cite{Case2023-an} in AmberTools23\cite{Li2016-vt}, based on Gaussian~16\cite{g16} calculations at the B3LYP/6-31G(d) level, and RESP\cite{Bayly1993-id} charges were adopted.
PFP calculations employed the \texttt{CRYSTAL\_U0\_PLUS\_D3} mode.
After geometry optimization using the L-BFGS algorithm (energy tolerance: \SI{10}{\kilo\joule\per\mole}), we performed a \SI{5}{\pico\second} MD run for equilibration, and the resulting structure was used as the initial configuration for subsequent simulations. 
Parameters of Langevin thermostat were a timestep of $\Delta t=\SI{0.5}{\femto\second}$ and a friction coefficient of $\SI{1}{\per\pico\second}$.
The initial configuration of the enzyme-substrate complex was prepared on the basis of the determined X-ray structure in Ref.~\citenum{Podgorski2022-qd} (PDB code: 7TP6) by manually placing the substrate.
We performed a \SI{3}{\nano\second} MD run for equilibration, and the resulting structure was used as the initial configuration for the subsequent well-tempered metadynamics simulation. 
A well-tempered metadynamics bias was applied to the CVs $(d_{\mathrm{OHC}},d_{\mathrm{COFe}})$ with a deposition stride of 500 MD steps, a Gaussian height of $\SI{0.2}{\electronvolt}\simeq \SI{4.6}{kcal\per\mol}$, Gaussian widths $\sigma_{d_{\mathrm{OHC}}}=\SI{0.25}{\angstrom}$ and $\sigma_{d_{\mathrm{COFe}}}=\SI{0.25}{\angstrom}$, and a bias factor of $\gamma=50.0$.
The initial configurations for umbrella sampling were selected from the preceding metadynamics trajectory by choosing the structures closest to the target reaction coordinates. 
Subsequently, a 100~ps MD simulation was performed for equilibration.
The umbrella sampling windows are summarized in Table~S1.
\newpage

\begin{longtable}{lrrr}
\caption{Umbrella sampling window parameters. $k$ is the force constant of the harmonic restraint.}\\
\hline
Window ID & $d_{\mathrm{OHC}}$ (\AA) & $d_{\mathrm{COFe}}$ (\AA)  & $k\;(\si{\electronvolt\per\angstrom\squared})$ \\
\hline
\endfirsthead
\hline
Window ID & $d_{\mathrm{OHC}}$ (\AA) & $d_{\mathrm{COFe}}$ (\AA)  & $k\;(\si{\electronvolt\per\angstrom\squared})$ \\
\hline
\endhead
US-000 & -1.7000 & -1.3500 & 10.0000 \\
US-001 & -1.4800 & -1.2400 & 10.0000 \\
US-002 & -1.2600 & -1.1300 & 10.0000 \\
US-003 & -1.0400 & -1.0200 & 10.0000 \\
US-004 & -0.8200 & -0.9100 & 10.0000 \\
US-005 & -0.6000 & -0.8000 & 10.0000 \\
US-006 & -0.3800 & -0.6900 & 10.0000 \\
US-007 & -0.1600 & -0.5800 & 10.0000 \\
US-008 & 0.0600 & -0.4700 & 10.0000 \\
US-009 & 0.2800 & -0.3600 & 10.0000 \\
US-010 & 0.5000 & -0.2500 & 10.0000 \\
US-011 & 0.5000 & -0.0750 & 10.0000 \\
US-012 & 0.5000 & 0.1000 & 10.0000 \\
US-013 & 0.5000 & 0.2750 & 10.0000 \\
US-014 & 0.5000 & 0.4500 & 10.0000 \\
US-015 & 0.5000 & 0.6250 & 10.0000 \\
US-016 & 0.5000 & 0.8000 & 10.0000 \\
US-017 & 0.5000 & 0.9750 & 10.0000 \\
US-018 & 0.5000 & 1.1500 & 10.0000 \\
US-019 & 0.5000 & 1.3250 & 10.0000 \\
US-020 & 0.5000 & 1.5000 & 10.0000 \\
US-021 & 0.2800 & -0.3600 & 20.0000 \\
US-022 & 0.3900 & -0.3050 & 20.0000 \\
US-023 & 0.5000 & -0.2500 & 20.0000 \\
US-024 & 0.5000 & -0.1625 & 20.0000 \\
US-025 & 0.5000 & -0.0750 & 20.0000 \\
US-026 & 0.5000 & -0.5000 & 20.0000 \\
US-027 & 0.5000 & -0.4000 & 20.0000 \\
US-028 & 0.5000 & -0.3000 & 20.0000 \\
US-029 & 0.5000 & -0.2000 & 20.0000 \\
US-030 & 0.5000 & -0.1000 & 20.0000 \\
US-031 & 0.5000 & 0.0000 & 20.0000 \\
US-032 & 0.5000 & 0.1000 & 20.0000 \\
US-033 & 0.5000 & 0.2000 & 20.0000 \\
US-034 & 0.5000 & 0.3000 & 20.0000 \\
US-035 & 0.5000 & 0.4000 & 20.0000 \\
US-036 & 0.5000 & 0.5000 & 20.0000 \\
US-037 & 0.6000 & -0.5000 & 20.0000 \\
US-038 & 0.6000 & -0.4000 & 20.0000 \\
US-039 & 0.6000 & -0.3000 & 20.0000 \\
US-040 & 0.6000 & -0.2000 & 20.0000 \\
US-041 & 0.6000 & -0.1000 & 20.0000 \\
US-042 & 0.6000 & 0.0000 & 20.0000 \\
US-043 & 0.6000 & 0.1000 & 20.0000 \\
US-044 & 0.6000 & 0.2000 & 20.0000 \\
US-045 & 0.6000 & 0.3000 & 20.0000 \\
US-046 & 0.6000 & 0.4000 & 20.0000 \\
US-047 & 0.6000 & 0.5000 & 20.0000 \\
US-048 & 0.7000 & -0.5000 & 20.0000 \\
US-049 & 0.7000 & -0.4000 & 20.0000 \\
US-050 & 0.7000 & -0.3000 & 20.0000 \\
US-051 & 0.7000 & -0.2000 & 20.0000 \\
US-052 & 0.7000 & -0.1000 & 20.0000 \\
US-053 & 0.7000 & 0.0000 & 20.0000 \\
US-054 & 0.7000 & 0.1000 & 20.0000 \\
US-055 & 0.7000 & 0.2000 & 20.0000 \\
US-056 & 0.7000 & 0.3000 & 20.0000 \\
US-057 & 0.7000 & 0.4000 & 20.0000 \\
US-058 & 0.7000 & 0.5000 & 20.0000 \\
US-059 & 0.8000 & -0.5000 & 20.0000 \\
US-060 & 0.8000 & -0.4000 & 20.0000 \\
US-061 & 0.8000 & -0.3000 & 20.0000 \\
US-062 & 0.8000 & -0.2000 & 20.0000 \\
US-063 & 0.8000 & -0.1000 & 20.0000 \\
US-064 & 0.8000 & 0.0000 & 20.0000 \\
US-065 & 0.8000 & 0.1000 & 20.0000 \\
US-066 & 0.8000 & 0.2000 & 20.0000 \\
US-067 & 0.8000 & 0.3000 & 20.0000 \\
US-068 & 0.8000 & 0.4000 & 20.0000 \\
US-069 & 0.8000 & 0.5000 & 20.0000 \\
US-070 & 0.9000 & -0.5000 & 20.0000 \\
US-071 & 0.9000 & -0.4000 & 20.0000 \\
US-072 & 0.9000 & -0.3000 & 20.0000 \\
US-073 & 0.9000 & -0.2000 & 20.0000 \\
US-074 & 0.9000 & -0.1000 & 20.0000 \\
US-075 & 0.9000 & 0.0000 & 20.0000 \\
US-076 & 0.9000 & 0.1000 & 20.0000 \\
US-077 & 0.9000 & 0.2000 & 20.0000 \\
US-078 & 0.9000 & 0.3000 & 20.0000 \\
US-079 & 0.9000 & 0.4000 & 20.0000 \\
US-080 & 0.9000 & 0.5000 & 20.0000 \\
US-081 & 1.0000 & -0.5000 & 20.0000 \\
US-082 & 1.0000 & -0.4000 & 20.0000 \\
US-083 & 1.0000 & -0.3000 & 20.0000 \\
US-084 & 1.0000 & -0.2000 & 20.0000 \\
US-085 & 1.0000 & -0.1000 & 20.0000 \\
US-086 & 1.0000 & 0.0000 & 20.0000 \\
US-087 & 1.0000 & 0.1000 & 20.0000 \\
US-088 & 1.0000 & 0.2000 & 20.0000 \\
US-089 & 1.0000 & 0.3000 & 20.0000 \\
US-090 & 1.0000 & 0.4000 & 20.0000 \\
US-091 & 1.0000 & 0.5000 & 20.0000 \\
US-092 & 1.1000 & -0.5000 & 20.0000 \\
US-093 & 1.1000 & -0.4000 & 20.0000 \\
US-094 & 1.1000 & -0.3000 & 20.0000 \\
US-095 & 1.1000 & -0.2000 & 20.0000 \\
US-096 & 1.1000 & -0.1000 & 20.0000 \\
US-097 & 1.1000 & 0.0000 & 20.0000 \\
US-098 & 1.1000 & 0.1000 & 20.0000 \\
US-099 & 1.1000 & 0.2000 & 20.0000 \\
US-100 & 1.1000 & 0.3000 & 20.0000 \\
US-101 & 1.1000 & 0.4000 & 20.0000 \\
US-102 & 1.1000 & 0.5000 & 20.0000 \\
US-103 & 1.2000 & -0.5000 & 20.0000 \\
US-104 & 1.2000 & -0.4000 & 20.0000 \\
US-105 & 1.2000 & -0.3000 & 20.0000 \\
US-106 & 1.2000 & -0.2000 & 20.0000 \\
US-107 & 1.2000 & -0.1000 & 20.0000 \\
US-108 & 1.2000 & 0.0000 & 20.0000 \\
US-109 & 1.2000 & 0.1000 & 20.0000 \\
US-110 & 1.2000 & 0.2000 & 20.0000 \\
US-111 & 1.2000 & 0.3000 & 20.0000 \\
US-112 & 1.2000 & 0.4000 & 20.0000 \\
US-113 & 1.2000 & 0.5000 & 20.0000 \\
US-114 & 1.3000 & -0.5000 & 20.0000 \\
US-115 & 1.3000 & -0.4000 & 20.0000 \\
US-116 & 1.3000 & -0.3000 & 20.0000 \\
US-117 & 1.3000 & -0.2000 & 20.0000 \\
US-118 & 1.3000 & -0.1000 & 20.0000 \\
US-119 & 1.3000 & 0.0000 & 20.0000 \\
US-120 & 1.3000 & 0.1000 & 20.0000 \\
US-121 & 1.3000 & 0.2000 & 20.0000 \\
US-122 & 1.3000 & 0.3000 & 20.0000 \\
US-123 & 1.3000 & 0.4000 & 20.0000 \\
US-124 & 1.3000 & 0.5000 & 20.0000 \\
US-125 & 1.4000 & -0.5000 & 20.0000 \\
US-126 & 1.4000 & -0.4000 & 20.0000 \\
US-127 & 1.4000 & -0.3000 & 20.0000 \\
US-128 & 1.4000 & -0.2000 & 20.0000 \\
US-129 & 1.4000 & -0.1000 & 20.0000 \\
US-130 & 1.4000 & 0.0000 & 20.0000 \\
US-131 & 1.4000 & 0.1000 & 20.0000 \\
US-132 & 1.4000 & 0.2000 & 20.0000 \\
US-133 & 1.4000 & 0.3000 & 20.0000 \\
US-134 & 1.4000 & 0.4000 & 20.0000 \\
US-135 & 1.4000 & 0.5000 & 20.0000 \\
US-136 & 1.5000 & -0.5000 & 20.0000 \\
US-137 & 1.5000 & -0.4000 & 20.0000 \\
US-138 & 1.5000 & -0.3000 & 20.0000 \\
US-139 & 1.5000 & -0.2000 & 20.0000 \\
US-140 & 1.5000 & -0.1000 & 20.0000 \\
US-141 & 1.5000 & 0.0000 & 20.0000 \\
US-142 & 1.5000 & 0.1000 & 20.0000 \\
US-143 & 1.5000 & 0.2000 & 20.0000 \\
US-144 & 1.5000 & 0.3000 & 20.0000 \\
US-145 & 1.5000 & 0.4000 & 20.0000 \\
US-146 & 1.5000 & 0.5000 & 20.0000 \\
\hline
\end{longtable}

\bibliography{ref}